\documentclass[sigconf, natbib=true, screen,nonacm,review=false,timestamp=false]{acmart}

\usepackage{enumitem}
\usepackage{balance}

\AtBeginDocument{%
  }

\copyrightyear{2025}
\acmYear{2025}
\setcopyright{cc}
\setcctype{by-nc-sa}
\acmConference[COMPASS '25]{ACM SIGCAS/SIGCHI Conference on Computing and Sustainable Societies}{July 22--25, 2025}{Toronto, ON, Canada}
\acmBooktitle{ACM SIGCAS/SIGCHI Conference on Computing and Sustainable Societies (COMPASS '25), July 22--25, 2025, Toronto, ON, Canada}
\acmDOI{10.1145/3715335.3736324}
\acmISBN{979-8-4007-1484-9/2025/07}

\begin{document}
\sloppy

\title[The Cloud Next Door]{The Cloud Next Door: Investigating the Environmental and Socioeconomic Strain of Datacenters on Local Communities}

\author{Wacuka Ngata}
\affiliation{%
  \institution{Massachusetts Institute of Technology}
  \city{}
  \country{}}

\author{Noman Bashir}
\affiliation{%
  \institution{Massachusetts Institute of Technology}
  \city{}
  \country{}}

\author{Michelle Westerlaken}
\affiliation{%
  \institution{Massachusetts Institute of Technology}
  \city{}
  \country{}}

\author{Laurent Liote}
\affiliation{%
  \institution{Massachusetts Institute of Technology}
  \city{}
  \country{}}

\author{Yasra Chandio}
\affiliation{%
  \institution{University of Massachusetts Amherst}
  \city{}
  \country{}}

\author{Elsa Olivetti}
\affiliation{%
  \institution{Massachusetts Institute of Technology}
  \city{}
  \country{}}

\renewcommand{\shortauthors}{Ngata et al.}

\begin{abstract}
Datacenters have become the backbone of modern digital infrastructure, powering the rapid rise of artificial intelligence and promising economic growth and technological progress. However, this expansion has brought growing tensions in the local communities where datacenters are already situated or being proposed. While the mainstream discourse often focuses on energy usage and carbon footprint of the computing sector at a global scale, the local socio-environmental consequences---such as health impacts, water usage, noise pollution, infrastructural strain, and economic burden---remain largely underexplored and poorly addressed. In this work\footnote{
\textbf{Acknowledgments.} We thank the reviewers for their thoughtful comments that helped in improving the quality of the paper. 
We note that our study was granted an IRB Exempt Status under the MIT IRB Protocol (Exempt ID: E-6721). This paper will appear at ACM COMPASS 2025 as a Late Breaking Work.
}, we surface these community-level consequences through a mixed-methods study that combines quantitative data with qualitative insights. Focusing on Northern Virginia's ``Data Center Valley,'' we highlight how datacenter growth reshapes local environments and everyday life, and examine the power dynamics that determine who benefits and who bears the costs. Our goal is to bring visibility to these impacts and prompt more equitable and informed decisions about the future of digital infrastructure.
\end{abstract}



\maketitle

\section{Introduction}
\label{sec:intro}
The ongoing digitization of society --- driven by the widespread adoption of digital devices and services --- has led to a steadily increasing demand for computing power. 
This trend is expected to accelerate with the rise of artificial intelligence (AI), whose tools are significantly more compute-intensive than traditional applications~\cite{Shehabi:2024:DCEnergy, Bhargs:2024:DCPower}. 
In response, major technology companies and AI stakeholders are rapidly building new datacenters to fuel this revolution~\cite{Smith:2025:GoldenAI, Reuters:2025:MetaWisconsin}. 
For instance, total datacenter capacity is projected to increase from 55GW in 2023 to between 171GW (low-range) and 298GW (upper-range) by 2030~\cite{Bhargs:2024:DCPower}. 

The unprecedented and unfettered growth in AI has led to significant concerns about the socio-environmental implications of AI development and use~\cite{Bashir:2024:Climate, Wu:2022:SustAI, Strubell:2020:Energy, Smith:2025:GoldenAI}. 
The socio-environmental implications are multifaceted and intersect with many sectors in our society through complex supply chains needed to produce AI hardware, construction of buildings near population hubs, added energy and power demand on the electric grid, and water use to cool down burning hot AI servers.
While datacenters' global energy, carbon, and water footprints have received growing attention, their localized impacts on surrounding communities remain largely overlooked. 
Understanding and addressing these local effects is essential to sustainably meeting the rising demand for datacenters.

Datacenters impact their local environments and surrounding communities in numerous ways. 
While datacenters are attractive to states and countries for the tax revenue they generate and to utility companies for their massive energy demand~\cite{Martin:2025:Extracting}, these benefits often do not extend to surrounding communities and certainly do not outweigh negative local impacts.
The widespread and adverse local impacts span multiple dimensions, including (i) environmental issues such as noise pollution~\cite{Monserrate2022EcologicalImpacts}, air pollution~\cite{Han2024UnpaidToll}, and water usage~\cite{Li2023WaterFootprintAI}, (ii) social impacts such as lack of amenities, increased blackouts, and lack of aesthetic appeals for datacenter buildings, and (iii) economic strains such as increased electricity costs~\cite{Martin:2025:Extracting}, and impact on life of household appliances~\cite{Bloomberg:2024:PowerQuality} (detailed in~\autoref{sec:methods}). 

In this ongoing effort, we aim to highlight and document the often-overlooked socio-environmental implications of datacenters on their surrounding communities. 
We aim to achieve this by employing both quantitative and qualitative methods, treating them as equally important in capturing the breadth and depth of local impacts of datacenters. 
Quantitative approaches help us measure and map the scale and distribution of these effects---such as changes in electricity costs, noise levels, or blackout frequency---while qualitative methods offer critical insight into lived experiences, decision-making dynamics, and community priorities that might otherwise remain invisible in a numerical data-driven study. 
Together, these approaches enable us to surface, contextualize, and communicate the full range of local concerns, allowing for more informed and just decisions about future datacenter development.

As a preliminary case study, we focus on a region where the rise of datacenters has visibly reshaped the physical and social landscape: Northern Virginia. West of Washington, D.C., this area---now home to over 450 facilities---hosts the world’s largest concentration of datacenters~\cite{TremaynePengelly2024DataCenterHubs}. 
A drive down Loudoun County Parkway reveals a patchwork of gated sites with datacenters, power substations, and backup generators, forming what is now commonly referred to as ``Data Center Valley.'' Directly bordering this valley are residential neighborhoods, where residents experience the everyday consequences of datacenter expansion. Through our combined qualitative and quantitative methods, we aim to capture and surface the realities of communities like these that are navigating the immediate and long-term impacts of digital infrastructure development.

In this work, we center on community experiences and deepen our understanding of datacenters' localized environmental and socioeconomic impacts. In doing so, we make three contributions.

\begin{enumerate}[leftmargin=*, topsep=0.1cm, itemsep=0.1cm]
\item We highlight the local impacts of datacenters, expanding the scope of current discourse beyond global energy and carbon metrics to include environmental, social, and economic dimensions experienced by surrounding communities.

\item We propose a methodological approach that combines quantitative and qualitative methods on equal footing to both document these local impacts and uncover the power dynamics that influence how and where datacenters are sited.
\item We instantiate this framework through a case study of Northern Virginia’s ``Data Center Valley,'' offering preliminary insights and a replicable model for studying data center-community interactions in other rapidly developing regions.
\end{enumerate}

\section{Related Work}
\label{sec:related}
Research on datacenters' environmental footprint has come from several academic disciplines, with key contributions from \emph{Critical Data Studies (CDS)}, \emph{Sustainable Human-Computer Interaction (HCI)}, and the \emph{Sustainable Computing} communities. 
Each field offers distinct perspectives on the role and impact of datacenter infrastructure~\cite{Zander2024FuturingCloud}. 
However, despite growing attention to environmental and societal implications, few efforts bridge these disciplinary silos to examine the localized, lived impacts of datacenters in communities where they are physically situated.

CDS frames digital infrastructure as a site of social, political, and ethical inquiry, drawing on critical theory, sociology, environmental studies, and science and technology studies to analyze the broader societal consequences of data infrastructures~\cite{Zander2024FuturingCloud}. A subset of this work focuses explicitly on datacenters, interrogating how they affect labor, land, and local governance structures~\cite{Edwards2023CriticalDataCenters}.
In parallel, Sustainable HCI research has increasingly moved beyond the interface to consider the environmental costs of computing infrastructure, emphasizing material resource use, such as energy, carbon, and water, linked to digital technologies. It employs a mix of quantitative and ethnographic methods to evaluate their broader implications~\cite{Zander2024FuturingCloud}. 
For instance, \citet{Li2023WaterFootprintAI} quantify the water footprint of AI workloads, revealing how environmental costs fluctuate with ambient temperatures and datacenter design. 
However, they emphasize the uncertainty and variability that complicate localizing these impacts. 
Similarly, \citet{Han2024UnpaidToll} estimate population-level health risks related to air pollution from diesel generators used in datacenter operations, linking them to elevated cardiovascular and respiratory illness risks.
WattTime, a major provider of carbon intensity data, has also released data on the health damage caused by the use of electricity at different locations using an intervention model for air pollution~\cite{Tessum:2017:INMAP, WattTime:2025:Health}.
However, neither study captures the fine-grained environmental impacts experienced at the neighborhood level or discusses socioeconomic factors.

The computer systems and computer architecture communities have long focused on optimizing the resource efficiency of datacenters by developing techniques to improve energy efficiency, workload scheduling, and resource management.
Recent work on sustainable computing has begun to directly explore carbon-aware system design~\cite{Bashir:2021:Enabling, Acun:2023:CarbonExplorer, Gupta:2022:ACT, Wang:2025:LowCarbonServer, Bashir:2023:HotAir, Wu:2022:SustAI} and  operation~\cite{ Souza:23:Ecovisor, Hanafy:23:CarbonScaler, Gsteiger:2024:Caribou}.
There has also been increasing interest in understanding the metrics for quantifying environmental impacts~\cite{Maji:2014:Mirage, Maji:2024:Untangling, Gandhi:2023:Metrics}. 
These studies typically operate at the scale of cloud platforms, proposing system-level optimizations and modeling aggregate impacts. 
While this work is crucial for making computing sustainable, it often abstracts away from the specific communities and geographies that host these large-scale computing infrastructures.

Together, CDS, HCI, and Sustainable Computing research have highlighted important aspects of datacenter infrastructure---its environmental costs, systemic efficiencies, and broader sociopolitical significance. 
However, these perspectives remain fragmented. In particular, there is a notable lack of empirical research that combines technical assessments with grounded, place-based inquiry into how datacenters affect the people and environments around them. Our work aims to address this gap by integrating qualitative and quantitative methods to document the local socio-environmental impacts of datacenters. We focus on communities adjacent to major datacenter regions and examine not just what the impacts are, but how they are perceived, negotiated, and distributed.

\section{Methodology and Early Outcomes}
\label{sec:methods}
Many of the global concerns associated with datacenter infrastructure---such as energy use, water stress, and air pollution---manifest in highly localized ways. 
Here, we focus on surfacing these effects at the community level, where datacenters are increasingly embedded within residential and semi-rural landscapes. 

\subsection{Identifying Local Impact Dimensions}
As an early outcome of our stakeholder engagement process (detailed in~\ref{sec:findings}), we have identified four key dimensions of local impact: environmental, social, economic, and infrastructural.

\begin{table*}[t] 
\centering 
\footnotesize 
\caption{Overview of high-level impact categories and relevant data sources} 
\vspace{-0.4cm} 
\begin{tabular}{p{1.6cm}p{2.8cm}p{2.8cm}p{2.8cm}p{2.8cm}p{2.8cm}} 
\toprule 
\textbf{Impact} & \textbf{Resource Strain} & \textbf{Power Grid Strain} & \textbf{Economic Effects} & \textbf{Noise Pollution} & \textbf{Air Pollution} \\ 
\midrule 
\textbf{Description} & 
Significant water and land usage & 
Branching effects of increased power demands & 
Increased electrical and water bills & 
Noise from diesel generators and IT equipment & 
Emissions from diesel generators and fossil-fuel-powered plants \\
 
\textbf{Metric} & 
Water Usage Effectiveness (WUE), land area covered, proximity to neighborhoods & 
Frequency and duration of power outages in surrounding areas & 
Changes in local electrical and water bills over time & 
Sound level measurements at various distances & 
Pollutant concentration levels (e.g., NO\textsubscript{x}, PM\textsubscript{2.5}) \\
 
\textbf{Unit of Impact} & 
L/kWh, acres, meters & 
\# of outages/year, avg. outage duration (minutes) & 
USD/month & 
Decibels (dB) & 
µg/m\textsuperscript{3} \\
 
\textbf{Data Sources} & 
Sustainability reports, public utility records, local zoning/GIS data & 
PJM Interconnection Data Miner, NOVEC outage maps, EIA grid reliability reports~\cite{PJM_DataMiner, NOVEC_StormCenter, EIA_ElectricityDataBrowser} & 
Local utility billing data, BLS average energy prices, EIA reports & 
Field measurements, dosimeters, NIOSH Sound Level Meter App, EPA guidelines~\cite{NIOSH_SLM_App} & 
EPA Air Quality System (AQS), NASA satellite data~\cite{EPA_AQS} \\
\bottomrule
\end{tabular}
\vspace{-0.4cm}
\label{tab:impact_summary}
\end{table*}

\smallskip
\noindent
\textbf{1 -- Environmental impacts} arise from both direct and indirect sources. 
Datacenters rely heavily on energy from large-scale power plants—facilities now increasingly co-located with datacenters to avoid delays in grid upgrades~\cite{WPR:2025:WisconsinDCs, Farney:2025:PowerPlay, Proctor:2025:Heart}. 
In addition to fossil-fuel-based generation, most datacenters maintain on-site diesel generators for backup power, contributing to localized air pollution~\cite{Han2024UnpaidToll}. 
Water use, both for cooling servers and for supporting power generation, places further strain on regional water reserves~\cite{Li2023WaterFootprintAI}. 
Noise pollution is also an emerging concern due to high-frequency sound produced by power distribution infrastructure, particularly in datacenters located near residential areas~\cite{Monserrate2022EcologicalImpacts}.

\smallskip
\noindent
\textbf{2 -- Social impacts} of datacenters are often framed only in terms of economic benefits, with claims about job creation and community development. 
However, datacenters do not create significant employment opportunities, as operations are increasingly automated and require minimal staffing. 
For example, in Northern Virginia, datacenters account for less than 0.5\% of the local workforce~\cite{USCensus_LoudounCounty, BLS_NorthernVirginiaEmployment}. 
Meanwhile, open land and green spaces near residential zones are being converted into datacenter campuses, displacing areas designated for community amenities and degrading quality of life~\cite{Upwind2024DataCenterGrowth}.

\smallskip
\noindent
\textbf{3 -- Economic impacts} are more complex. While datacenters contribute substantial tax revenue to local governments and utility providers, these financial benefits are often not equitably distributed. As energy-intensive infrastructure, datacenters drive up local power demand, which requires costly grid upgrades. These costs are frequently passed on to residential customers. 
For example, datacenter growth in Northern Virginia is expected to increase monthly residential electricity bills by \$37 by 2040~\cite{JLARC2024DataCentersVirginia}.

\smallskip
\noindent
\textbf{4 -- Infrastructural strain} is an emerging area of concern. As datacenter demand surges, local grids face new stresses. In response, extreme proposals—such as the construction of modular nuclear reactors—have been floated to meet growing energy needs~\cite{UtilityDive2023DataCenterNuclear}. In the meantime, insufficient capacity can result in degraded power quality or frequent blackouts in nearby residential neighborhoods.

\subsection{Assessing Impacts using Mixed-Methods}
To study these impact areas in context, we propose adopting a mixed-methods approach that combines quantitative measurement of environmental and economic effects with qualitative analysis of stakeholder perspectives and local decision-making processes.

\smallskip
\noindent
\textbf{Quantitative Approach.}
To capture the measurable impacts of datacenters at the local level, we have compiled metrics and data sources corresponding to five high-level concerns: resource strain, power grid stress, economic costs, noise pollution, and air pollution. Table~\ref{tab:impact_summary} summarizes these impact categories alongside representative metrics and data sources. 
These include water use effectiveness, land coverage, outage frequency, utility billing records, decibel levels, pollutant concentrations, public utility databases, environmental monitoring systems, and near-site measurements.

\smallskip
\noindent
\textbf{Qualitative Approach.}
To complement the quantitative assessment, we plan to conduct qualitative research to understand how datacenter impacts are perceived and negotiated within local communities. Our objectives are threefold. First, we aim to uncover effects not captured by environmental metrics or utility data—those revealed through stakeholder narratives and community experience. Second, we seek to contextualize the quantitative findings by linking them to the lived realities of affected populations. Third, we aim to identify the socio-political structures and decision-making dynamics that influence the practical development of datacenters.

We are collecting qualitative data through document analysis (e.g., local newsletters, public reports), participation in community events, engagement with stakeholder groups on social media, and semi-structured interviews. This ongoing research has received IRB exempt clearance under the MIT protocol (IRB Exempt ID E-6721). Interviews and other forms of engagement are being conducted only with the participant's consent, and all collected data is being anonymized unless otherwise specified by the participant.

Interviewees include a range of stakeholders such as local residents, utility representatives, planners, and public officials. The initial findings are detailed in~\autoref{sec:findings}. 
Among residents, we have spoken with small business owners and retirees, many of whom have relocated from other eastern states to suburban and semi-rural Virginia. These individuals help reflect the everyday concerns and lived experiences of communities near datacenters. 
We also include participants affiliated with environmental organizations, including professionals in national park conservation, who bring valuable insight into the environmental impacts of datacenter development. 
On the industry side, interviewees include engineers with experience at utility companies, such as Dominion Energy, as well as planners and managers contracted by large cloud service providers, including Google. 
These participants offer perspectives on how utilities and datacenter operators respond to community concerns. All interviews will be transcribed and analyzed thematically, with a focus on how the physical footprint of datacenters is experienced and contested by different groups.

\smallskip
\noindent
\textbf{Integrating Quantitative and Qualitative Findings.}
By combining these two approaches, we aim to produce a multi-dimensional view of datacenter infrastructure that accounts for measurable outcomes and their social realities. Environmental data provides indicators of resource strain and infrastructural pressure, while qualitative insights highlight how those indicators translate into everyday impacts and political friction. 
The integration of both forms of data allows us to move beyond abstract metrics and toward a grounded understanding of who bears the cost of digital expansion—and how those costs are justified, resisted, or reimagined.

\subsection{Preliminary Findings}
\label{sec:findings}
We next outline our preliminary quantitative and qualitative findings from Northern Virginia's Datacenter Valley case study.

\noindent
\textbf{Quantitative Outcomes.}
We conducted preliminary noise tests in Ashburn, VA, comparing ambient levels in a control neighborhood two miles from a datacenter to a residential area just 200 feet away, using the NIOSH Sound Level Meter app~\cite{CDC:2025:NIOSH}. As shown in Table~\ref{tab:noise_levels}, noise was consistently higher near datacenters, likely due to persistent operational sounds. 
This low-frequency hum may affect local well-being and adds to the datacenters' environmental footprint. 
We plan to expand the study to more sites and measurements.

To understand the effect on power quality, we analyzed the data from a recent study conducted by Bloomberg~\cite{Bloomberg:2024:PowerQuality}.
In Loudoun County, Virginia, more than 6.8\% of the homes experienced atleast one monthly reading exceeding the 8\% total harmonic distortion (THD) that can damage appliances. 
In Prince William County, more than two dozen out of the 1100 deployed sensors experienced upto 13\% THD. 
These distortions are strongly linked to proximity to datacenters and tend to peak at night when datacenters are the primary consumers of the electric grid power. 

We are currently analyzing utility providers' outage frequency and load distribution data to assess whether datacenter expansion correlates with power quality issues in adjacent neighborhoods. This analysis is ongoing, and we are developing additional data pipelines to track spatiotemporal load trends and outage reports over time and across geographical locations. Thus far, these quantitative data points appear to corroborate the infrastructural strain that has concerned local communities.

\begin{table}[t] 
\centering 
\caption{Average noise levels measured at varying distances from the datacenter site in the afternoon.} 
\vspace{-0.3cm} 
\begin{tabular}{p{4cm}p{3.5cm}} 
\toprule 
\textbf{Location} & \textbf{TWA Noise Level [dB]} \\ 
\midrule 
Neighborhood 2 miles away & 22.3 \\ 
Neighborhood 200 feet away & 28.0 \\ 
\bottomrule 
\end{tabular} 
\label{tab:noise_levels} 
\vspace{-0.7cm}
\end{table}


\smallskip
\noindent
\textbf{Qualitative Outcomes.}
Through our early engagements to gather stakeholder perspectives, we have identified three dominant attitudes toward datacenters: \textit{concerned}, \textit{incentivized}, and \textit{indifferent}. These perspectives vary across ten types of stakeholders, including residents, local governments, utility companies, and industry operators. Stakeholders differ widely in their awareness of datacenter impacts, influence over development processes, and perceived benefits or harms. Mapping these relationships is helping us surface the complex power dynamics and competing narratives that shape local responses to increasingly ubiquitous digital infrastructure.

Thus far, we have received responses and engaged in discussions with individuals from environmental organizations and local communities, who have made up the majority of our initial participant pool. 
These early conversations have raised various concerns that help us better capture the social impacts of datacenters. 
For instance, from these early conversations, local residents have voiced their surprise at the rapid transformation brought about by the datacenter development, often describing a sense of hopelessness in their efforts to retain the quality of life they sought when moving into the area. 
As part of our ongoing research, we intend to engage in further discussions, soliciting the viewpoints of a broader range of stakeholders to obtain valuable insights into the potential impact of datacenters on not only the physical landscape but also the perception and relationship of residents with their surroundings. 

\section{Discussion And Future Work}
\label{sec:discussion}
By combining qualitative insights with quantitative data, our goal is to develop a comprehensive database that enhances both the capability for impact evaluation of existing datacenters and the site selection process for future facilities. This resource is intended to support more informed decision-making around project approvals, rejections, and sustainability appraisals.

To build a robust dataset, we will expand our conversations to ensure we engage with a broad range of stakeholders, with a particular emphasis on those directly affiliated with datacenter development and operations. In addition, we will continue conducting near-site experiments (e.g., measuring ambient noise levels near facilities) and acquiring data related to the local power grid dynamics. As we collect data across these varied sources, we will apply thematic analysis to uncover patterns and relationships, contributing detailed and meaning-rich data points to the overall dataset. 

To ensure that our dataset is robust and generalizable, we will iteratively apply our methods across different types of datacenters. Given that Northern Virginia is the hub of hyperscale datacenters, our initial analysis focuses on the ecological effects of this specific infrastructure model. In the future, we will assess other locations and datacenter types, enabling comparative evaluations across infrastructure designs and operational models.
Beyond informing public policy and planning decisions, this publicly accessible data repository can also serve as a catalyst for industry change. By surfacing localized impacts and providing actionable insights, we aim to motivate vendors, operators, and major cloud service providers to adopt more sustainable and community-aware practices.

A key challenge in this work is obtaining accurate, granular data on datacenters’ water usage and electrical loads, including the volumes consumed and the specific sources of these resources. Due to the limited availability of publicly reported data, our quantitative analyses will, in some cases, rely on well-founded estimations to assess the strain placed on local systems.
While estimations introduce uncertainty, they do not diminish the value of the overall analysis. Rather, the qualitative insights collected from a broad spectrum of stakeholders will play a critical role in contextualizing these metrics, revealing overlooked themes, and highlighting where decision-making power resides. Together, these insights will guide the development of a meaningful, actionable database that supports both policy interventions and industry accountability.

\section{Conclusion}
\label{sec:conclusion}
With this late-breaking work, we aim to bridge a critical gap in understanding the collective ecological impacts of datacenters on their immediate environments by using both quantitative and qualitative approaches. We aim to address the following key questions: What are the diverse effects of datacenters on local ecosystems? How can these impacts be effectively measured? And who are the stakeholders best positioned to respond to these findings? Answering these questions is essential for enabling more informed and equitable decision-making around data infrastructure development at the local level. By offering a clearer view of the physical and socio-political consequences of datacenter expansion, this work contributes to advancing digital infrastructure in ways that center sustainability and community well-being.

\balance
\bibliographystyle{ACM-Reference-Format}
\bibliography{paper}


\end{document}